\documentclass[reprint,prx,amsmath,amssymb,aps,superscriptaddress,longbibliography]{revtex4-2}

\usepackage{color,soul}
\usepackage{graphicx}
\usepackage{dcolumn}
\usepackage{bm}
\usepackage{hyperref}
\usepackage{mathtools}
\hypersetup{colorlinks=true, linktoc=all, linkcolor=blue, linktocpage, citecolor=blue}

\newcommand{\bs}{\boldsymbol}

\begin{document}

\title{Ensemble inequivalence in the design of mixtures with super-Gibbs phase coexistence}

\author{Filipe C. Thewes}
\email{filipe.thewes@ds.mpg.de}
\affiliation{%
 Institut für Theoretische Physik, Georg-August-Universität Göttingen, 37077 Göttingen, Germany
}
\affiliation{Max Planck Institute for Dynamics and Self-Organization, Am Faßberg 17, 37077 Göttingen, Germany\\}%
\author{Peter Sollich}%
\affiliation{%
 Institut für Theoretische Physik, Georg-August-Universität Göttingen, 37077 Göttingen, Germany
}%
\affiliation{King's College London, Department of Mathematics, Strand, London WC2R 2LS, U.K.\\}

\date{\today}

\begin{abstract}
Designing the phase behavior of multicomponent mixtures is a rich area with many potential applications. One key question is how more than $M+1$ phases, as would normally be allowed by Gibbs' phase rule at generic temperature in a mixture of $M$ molecular species, can be made to coexist in equilibrium. In the grandcanonical ensemble, such super-Gibbs phase equilibria can be realized by tuning the interactions among the $M$ species. This introduces $\sim M^2$ additional degrees of freedom and hence a superlinear  number of phases that can coexist. We show that, surprisingly, there is no straightforward equivalence to the situation in the experimentally relevant canonical ensemble: here only a subset of the grandcanonical phases will generically be realized. This subset is determined by interfacial tensions in addition to bulk free energies.
Using a graph-theoretical approach, we determine a sufficient set of inequalities for the interfacial tensions for which all grandcanonical phases are realized so that equivalence of ensembles is effectively restored. We illustrate the design method for a two-component mixture with four coexisting phases and point out the route for generalizing this to a higher number of components.
\end{abstract}

\maketitle

\section{Introduction}

Multicomponent mixtures may demix via liquid-liquid phase separation (LLPS) into multiple phases, possibly showing a number of metastable states~\cite{Meunier2000,Zwicker2022,Mao2019}. For example, a binary system (e.g.\ water and oil) will remain mixed or split into water-rich and oil-rich domains at a given pressure, temperature, and chemical potentials of water and oil molecules~\cite{Sollich2001,Bray1994}. Changing these parameters changes the equilibrium states of the mixture and a protocol can be used to engineer a desired set of macroscopic properties~\cite{Mao2020,Jacobs2021}. For binary systems this is relatively straightforward, but becomes a much richer problem in multi-component mixtures. In contrast to their single component counterparts, such systems present very complex behavior, which makes it hard to engineer a chosen set of properties~\cite{Mao2020,Jacobs2021,Cho2023,peters_2021}. The resulting mixture properties due to LLPS are not only widely used in the materials sciences~\cite{Haase2014,Feric2016,Simon2017,Lu2020,Kaur2021} and food industry~\cite{Mathijssen2023}, but have also been shown to be a key driver in the development of protocells, in which LLPS allows for the formation of stable membraneless organelles such as the nucleus~\cite{Mao2020,Zwicker2022,Jacobs2021,Cho2023}. 

Surprisingly, using a random graph approach, it was shown recently~\cite{Jacobs2021} that mixtures with a large number $M$ of components and {\em designed} interactions can assemble into $n\sim O(M^2)$ phases, apparently contradicting the well established Gibbs' phase rule, which predicts a maximum of $M+1$ phases at a generic temperature. Later it was argued that the apparent paradox of such ``super-Gibbs'' phase splits was due to the existence of linearly dependent designed coexisting phases~\cite{Cho2023}. From an equation of state perspective, multiphase coexistence beyond Gibbs' phase rule has also been observed in mixtures of colloids and polymers~\cite{Vega1997,Akahane2016,peters_2020}, once the shape of the constituent particles is allowed to be fine-tuned.

In this work we show that, while designed super-Gibbs phase coexistence in the grandcanonical ensemble can be understood from a straightforward extension of the standard Gibbs phase rule that includes design parameters as additional degrees of freedom, the same is not true in the experimentally more relevant canonical ensemble where particle numbers are fixed: here not all grandcanonically designed phases will coexist, unless additional conditions are satisfied.

The importance of this result stems from the fact that equivalence of ensembles~\cite{Georgii1995} is one of the key concepts of statistical physics; it states that the constraints imposed by the choice of ensemble become irrelevant in the thermodynamic limit. This means that different ensembles yield the same equilibrium macroscopic properties once the system is taken to be large enough. For liquid mixtures, one of these macroscopic properties is the phase split, which captures which phases are observed for a given set of intensive parameters. We will show that for designed super-Gibbs phase coexistence in the canonical ensemble, the interfacial properties remain relevant even in the thermodynamic limit: they determine which phases will coexist, resulting in an inequivalence with the grandcanonical ensemble (Sec.~\ref{sec:canonical}). We then demonstrate that under suitable conditions on the interfacial tensions between different phases -- which are in addition to the tuning of bulk free energies required in the grandcanonical ensemble -- designed super-Gibbs phase coexistence can be realized also in the canonical ensemble, thus effectively restoring ensemble equivalence. We give sufficient conditions for this for a generic number of mixture components in Sec.~\ref{subsec:multiSpec}. We finally discuss the problem of interfacial tension design and point out a route to meeting the previously derived conditions via a design protocol (Sec.~\ref{sec:designInterf}).

\section{Gibbs' Phase Rule in the grandcanonical ensemble}
Gibbs' phase rule can be derived by considering a mixture with $M$ components or species, with fixed total volume $V$ and temperature $T$. Each of $n$ coexisting bulk phases is then specified by its composition, i.e.\ by the component densities $\rho_i=N_i/V$ for $i=1,\ldots,M$. Chemical and mechanical equilibrium require that the pressure $P$ and chemical potentials $\mu_i$ are the same across all phases. This gives $(n-1)(M+1)$ constraints on $nM$ variables, leaving a total of  $M+1-n$ degrees of freedom. The maximum number of phases that can coexist is then $n=M+1$.

The reasoning above, which uses the canonical ensemble, can be rephrased in the grandcanonical ensemble, where the generalization to designed interactions is straightforward. One imposes, in addition to temperature $T$ and volume $V$ as before, the chemical potential $\mu_i$ of each mixture component, while the particle numbers $N_i$ and hence densities $\rho_i=N_i/V$ become fluctuating variables. 
The probability of seeing a certain overall composition 
$\boldsymbol{\rho}=(\rho_1,\ldots,\rho_M)$ is then 
\begin{equation}
p(\boldsymbol{\rho})\propto e^{-\beta V[f(\boldsymbol{\rho})-\boldsymbol{\mu}^\mathsf{T}\boldsymbol{\rho}]}
\end{equation}
where $\boldsymbol{\mu}^\mathsf{T}\boldsymbol{\rho}=\sum_i \mu_i \rho_i$, $f(\boldsymbol{\rho})$ is the free energy density for the given composition and $\beta=1/(k_{\rm B}T)$ is the inverse temperature.
Each coexisting bulk phase $\alpha$ corresponds to a peak in the distribution $p(\boldsymbol{\rho})$, of width $\propto V^{-1/2}$. Denoting the region containing this peak by $R_\alpha$, the phase coexistence condition is that the probability of seeing each phase is equal, 
\begin{equation}
    \int_{R_1} d\boldsymbol{\rho}\, p(\boldsymbol{\rho}) = \ldots = \int_{R_n} d\boldsymbol{\rho}\, p(\boldsymbol{\rho}).
    \label{equalProb}
\end{equation}
For large $V$ each peak can be taken as Gaussian, so that 
\begin{equation}
    \int_{R_\alpha}d\boldsymbol{\rho}\, p(\boldsymbol{\rho}) \propto c_\alpha V^{-M/2}e^{-\beta V(f^{(\alpha)}-\boldsymbol{\mu}^\mathsf{T}\boldsymbol{\rho}^{(\alpha)})}
    \label{prob0}
\end{equation}
up to a proportionality factor that is common to all phases. Here 
$\boldsymbol{\rho}^{(\alpha)}$ is the composition of phase $\alpha$ as given by the position of the maximum of the peak, $f^{(\alpha)}=f(\boldsymbol{\rho}^{(\alpha)})$ is the corresponding free energy density, and $c_\alpha$ can be expressed in terms of the curvature of $f(\boldsymbol{\rho})$ at the peak. Inserting into~\eqref{equalProb} shows that
\begin{equation}
f^{(\alpha)}-\boldsymbol{\mu}^\mathsf{T}\boldsymbol{\rho}^{(\alpha)}-\frac{T}{V}\ln c_\alpha
\end{equation}
has to be equal in all phases. The last term can be ignored in the thermodynamic limit (for finite $V$, it gives $1/V$-corrections to the values of the chemical potentials at coexistence) so that one retrieves the condition of equal pressures 
\begin{equation}
P^{(\alpha)} = -f^{(\alpha)}+\boldsymbol{\mu}^\mathsf{T}\boldsymbol{\rho}^{(\alpha)}
\label{prob}
\end{equation}
at coexistence. When $n$ phases coexist, this gives $n-1$ conditions on the $M$ chemical potentials $\boldsymbol{\mu}$, leaving $M-n+1$ degrees of freedom. {Consistent with the idea of equivalence of ensembles, this conclusion is the same as in the canonical ensemble.}

The above argument now generalizes directly to the design of interactions. Each adjustable parameter in the bulk free energy $f(\boldsymbol{\rho})$, for example an interaction strength between two species $i$ and $j$, provides one additional variable that can be tuned in addition to the $M$ chemical potentials. Thus, if there are $D$ design parameters, we have $M+D$ parameters overall. With $n-1$ equal pressure constraints as before, this leaves
\begin{equation}
M+D-n+1
\label{Gibbs_extended}
\end{equation}
degrees of freedom, giving a maximum number $M+D+1$ of coexisting phases at generic temperature. The design or fine-tuning of pairwise interactions, for example, provides $M(M+1)/2$ tunable parameters, allowing for a maximum of $n\sim M^2$ phases. This scaling is consistent with an upper bound on $n$ for pairwise interaction design in~\cite{Jacobs2021}: their graph theoretical approach yields a maximum value $n\sim M^{(2s-1)/(s+1)}<M^2$ where $s$ is the number of enriched species in each phase. 

A simpler example is provided by triple point coexistence in single-component systems, such as ice--water--vapour: the temperature tuning required to reach such a triple point is effectively the same as tuning an interaction strength $\epsilon$ at fixed $T$, given that all free energies and chemical potentials (in units of $k_{\rm B}T$) can only depend on $T/\epsilon$. Designing intermolecular interactions in mixtures with $M>1$ species has the analogous effect. 

In~\cite{Vega1997}, Vega and Monson proposed an extension of Gibbs' phase rule similar to (\ref{Gibbs_extended}) for mixtures of colloidal systems with tunable length scales, i.e.\ their relative shapes, which was confirmed later by Monte Carlo simulations~\cite{Akahane2016}. An extended rule for fine-tuned particle-based phase coexistence was later introduced by Peters et al~\cite{peters_2020} to {include degrees of freedom associated with microscopic length-scales}. Note that if, beyond pairwise interactions, also three-body potentials can be tuned, the number of coexisting phases can in principle be pushed to $n\sim M^3$, and so on for higher-order interactions.

\section{Designed Super-Gibbs Phase Coexistence in the Canonical Ensemble}
\label{sec:canonical}

It is important to note that in the grandcanonical reasoning above, configurations containing multiple phases in the same volume $V$ do not play any role. This is because the resulting interfaces will incur a free energy cost scaling as $V^{(d-1)/d}$ in $d$ dimensions. This leads to 
any configuration containing interfaces being exponentially suppressed for large $V$, compared to the dominant pure bulk phase configurations. Thus effectively only the latter configurations are observable and contribute to the probabilities (\ref{equalProb}) of each phase.

\begin{figure}[hbt]
\includegraphics[width=\columnwidth]{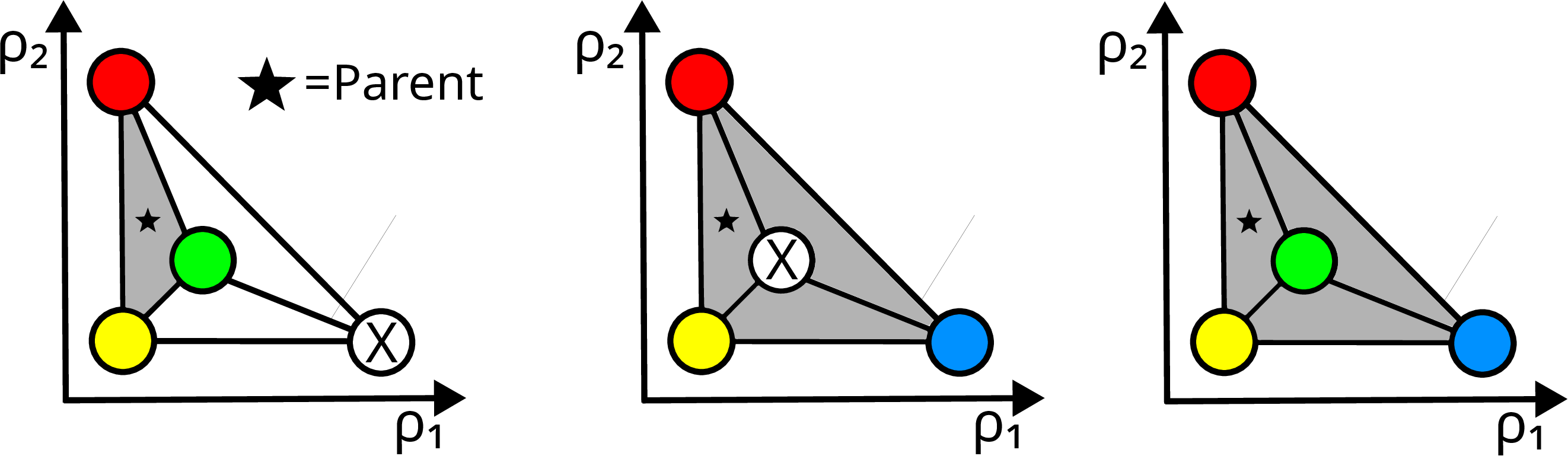}
\caption{Illustrative example with $M=2$ species and $m=4$ grandcanonically designed phases (circles). In the canonical ensemble, the lever rule can be satisfied as a convex linear combination of either three (left and middle) or four (right) phases. Out of these options, the phase split that will be observed at equilibrium is the one that minimizes the interfacial free energy, which also depends on the spatial arrangement of the phases. }
\label{sketchLever}
\end{figure}

In the canonical ensemble, i.e.\ in the typical experimentally relevant setting 
of a fixed total number of particles of each species, or equivalently fixed ``parent'' composition $\boldsymbol{\rho}^{(0)}$~\cite{Sollich2001}, interfaces must necessarily occur, and the argument above must be extended to include their effects. {If a system prepared with a given parent composition splits into multiple phases then, 
by particle number conservation, 
the average composition of these coexisting phases, weighted by the fractions $v_\alpha$ of system volume that they occupy, must equal the parent composition}:
\begin{equation}
    \boldsymbol\rho^{(0)} = \sum_\alpha v_\alpha \boldsymbol\rho^{(\alpha)}\quad \textrm{with}\quad \sum_\alpha v_\alpha=1.
    \label{lever}
\end{equation}
In the following it will be useful to think of this ``lever rule''~\cite{Sollich2001} constraint in geometric terms, i.e.\ in the $M$-dimensional space of all possible compositions $\boldsymbol{\rho}$. The r.h.s.\ of (\ref{lever}) is a convex linear combination of the $\boldsymbol\rho^{(\alpha)}$, i.e.~one with non-negative coefficients that add up to $1$. The set of all such convex linear combinations defines the convex hull of the vectors $\boldsymbol\rho^{(\alpha)}$. This phase coexistence region is a convex polyhedron; for a two-species system ($M=2$) with $M+1=3$ coexisting phases it reduces to the familiar three-phase triangle in a two-dimensional ($\rho_1,\rho_2$) phase diagram.
More generally, barring degeneracies or symmetries that make the vectors $\boldsymbol\rho^{(\alpha)}$ linearly dependent, their convex hull is a polyhedron with a nonzero ($M$-dimensional) volume (see Fig.~\ref{sketchLever}). 

Now if there are $M+1$ coexisting phases, then any parent composition $\boldsymbol\rho^{(0)}$ within the phase coexistence region can be represented uniquely as a convex linear combination of the $ \boldsymbol\rho^{(\alpha)}$.
If, on the other hand, a larger number of phases have been designed to coexist in the grandcanonical ensemble, then distinct sets of $M+1$ or more phases will be able to recover the parent in this way (Fig.~\ref{sketchLever}), and out of these potential phase splits the one that is realized at (canonical) equilibrium will be the one with the lowest free energy. 
However, the bulk free energy contributions are no help here as they are the same for all possible phase splits by design: from (\ref{prob}) and using the coexistence conditions $P^{(\alpha)}=P$ one has for the total free energy
\begin{equation}
V\sum_\alpha v_\alpha f^{(\alpha)} = V\sum_\alpha v_\alpha (\boldsymbol{\mu}^\mathsf{T}\boldsymbol{\rho}^{(\alpha)}-P) = V
(\boldsymbol{\mu}^\mathsf{T}\boldsymbol{\rho}^{(0)}-P)
\label{bulk}
\end{equation}
for any set of $v_\alpha$ satisfying the lever rule (\ref{lever}).
Interfacial terms not included in (\ref{bulk}) therefore become the determining factor that governs which phase split is realized. {This breaks the equivalence between the canonical and grandcanonical ensembles}. We will proceed to show that, indeed, it is the interfacial properties that determine the coexisting phases in canonical equilibrium, so that designing coexistence of more than $M+1$ phases requires the interfacial tensions between the different bulk phases to satisfy specific inequalities.

\subsection{Hamiltonian description: periodic solutions}
To include interfacial terms in our considerations, we consider an arbitrary spatially varying composition $\boldsymbol{\rho}(\boldsymbol{r})$ and  write the total free energy of the system as 
\begin{equation}
    F = \int d\boldsymbol r \left [f(\boldsymbol{\rho}) + \frac{1}{2}\nabla\boldsymbol{\rho}^{\mathsf{T}}\boldsymbol{K}(\boldsymbol \rho)\nabla\boldsymbol{\rho} \right ] 
    \label{freeEn0}
\end{equation}
where as before $f$ defines the bulk free energy of a locally homogeneous system with densities $\boldsymbol\rho=(\rho_i)_{i=1\ldots M}$, while the positive definite matrix $\boldsymbol K$ captures the interfacial free energy contributions. Here and below we use $\boldsymbol a^\mathsf{T}\boldsymbol b$ for the scalar product in the $M$-dimensional species space while the dot ($\cdot$) notation is used for dot products in $d$-dimensional Euclidean space. Using this convention, the interfacial free energy density can be written more explicitly as $\frac{1}{2}\nabla\boldsymbol{\rho}^{\mathsf{T}}\boldsymbol{K}(\boldsymbol \rho)\nabla\boldsymbol{\rho}=\frac{1}{2}\sum_{ij} K_{ij}\nabla\rho_i\cdot\nabla \rho_j$.

For a generically $\boldsymbol\rho$-dependent matrix $\boldsymbol K\equiv \boldsymbol K(\boldsymbol \rho)$ of interfacial coefficients, the chemical potentials follow by taking the functional derivative of~\eqref{freeEn0}, i.e.\
\begin{equation}
\begin{split}
    \frac{\delta F}{\delta \rho_i} =\mu_i = &\frac{\partial f}{\partial \rho_i}  - \sum_j K_{ij}\nabla^2\rho_j\\
    & + \frac{1}{2}\sum_{j,l}\nabla\rho_j\cdot\nabla\rho_l\left [\frac{\partial K_{jl}}{\partial\rho_i} - 2\frac{\partial K_{ij}}{\partial\rho_l} \right ]
    \end{split}
    \label{chemPot0}
\end{equation}
At equilibrium, each $\mu_i$ must be constant across the entire system. 
This condition can be seen physically as coming from equilibrium under particle exchange between local regions of the system, or more mathematically from the requirement of minimizing the total free energy $F$, subject to the particle number conservation constraint $\int\,d\boldsymbol{r}\boldsymbol{\rho}(\boldsymbol{r})  = V\boldsymbol{\rho}^{(0)}$ and with the $\mu_i$ being the appropriate Lagrange multipliers.

We note for later that the interfacial free energy contribution to the total free energy can be determined by subtracting~\eqref{bulk} from~\eqref{freeEn0}. Bearing in mind the number conservation constraint, the result can be written in the form
\begin{equation}
 \Delta F = \int d\boldsymbol r \left [V_{\rm eff}(\boldsymbol{\rho}) 
 + \frac{1}{2}\nabla\boldsymbol{\rho}^{\mathsf{T}}\boldsymbol{K}(\boldsymbol \rho)\nabla\boldsymbol{\rho} \right ] 
    \label{F_interface}
\end{equation}
with 
\begin{equation}
    V_{\rm{eff}}(\boldsymbol\rho) = f(\boldsymbol\rho)-\boldsymbol\rho^\mathsf{T}\boldsymbol\mu + P
\end{equation}
By construction, each (grandcanonically designed) phase $\boldsymbol{\rho}^{(\alpha)}$ is a global minimum of $V_{\rm eff}(\boldsymbol{\rho})$, with $V_{\rm{eff}}(\boldsymbol{\rho}^{(\alpha)})=0$. The common tangent plane that geometrically embodies the phase coexistence conditions is simply $V_{\rm eff}=0$, so that $V_{\rm eff}(\boldsymbol{\rho})$ can also be thought as a tangent plane distance~\cite{Sollich1998, Sollich2001}.

\begin{figure}[hbt]
\includegraphics[width=\columnwidth]{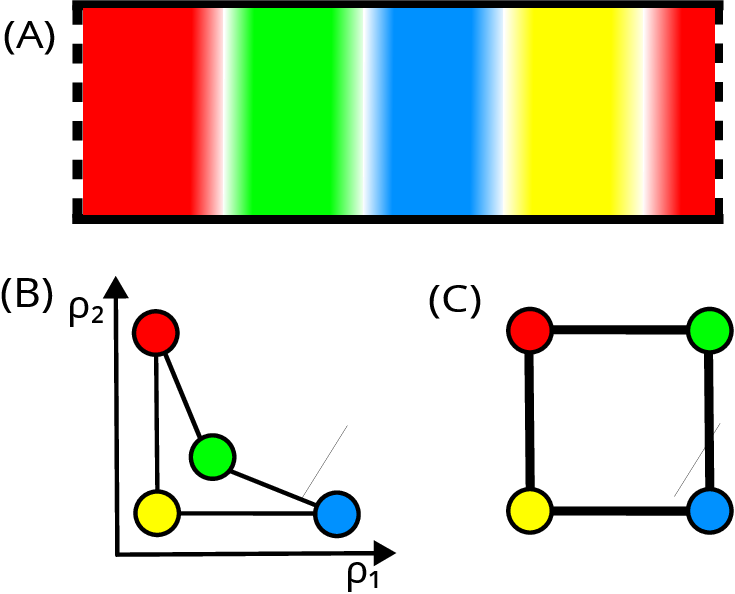}
\caption{Two species with four coexisting phases. (A) spatial arrangement of phases in a slab geometry and with periodic boundary conditions. (B)  Representation of the arrangement (A) in the $(\rho_1,\rho_2)$-plane, with circles positioned at the coexisting phases $\boldsymbol\rho^{(\alpha)}$ and interfaces as connecting lines. (C) Abstract graph representation of (A,B), showing  phases as nodes and interfaces as links.}
\label{sketchSlab}
\end{figure}

We return now to the spatial arrangement of  coexisting phases at equilibrium, which has to be obtained by solving~\eqref{chemPot0} (with constant $\mu_i$) for $\boldsymbol{\rho}(\boldsymbol{r})$. Importantly, in one spatial dimension this problem can be mapped to one in Hamiltonian mechanics. Calling the spatial coordinate $x$ so that $\nabla \to d/dx$, one can multiply~\eqref{chemPot0} by $d\rho_i/dx$ and sum over $i$ to find~\cite{Bray1994}
\newcommand{\ham}{\mathcal{H}}
\newcommand{\lagr}{\mathcal{L}}
\begin{equation}    
    \frac{d\ham}{dx}=0, \qquad 
    \ham = \frac{1}{2}\frac{d\boldsymbol\rho^{\mathsf{T}} }{dx}\boldsymbol K \frac{d\boldsymbol\rho}{dx}  -V_{\rm eff}(\boldsymbol{\rho}) 
    \label{hamiltonian_z}
\end{equation}    
This can be viewed as a conservation law for a Hamiltonian $\ham$: $x$ plays the role of time, $\boldsymbol \rho$ and $d\boldsymbol\rho/dx$ are the generalized coordinates and velocities, respectively,
and $\boldsymbol K$ corresponds to a mass tensor defining the effective kinetic energy. The same analogy also holds in higher spatial dimensions provided that the $\rho_i$ are functions of only a single spatial coordinate $x$, which physically means that the coexisting phases are arranged next to each other in a ``slab'' geometry with flat interfaces between them as shown in Fig.~\ref{sketchSlab}. (For other geometries, {a stress-tensor type of conservation law still exists}~\cite{Dean2009,Krger2018}.) We focus in this paper on the slab geometry case, but return in the concluding section to the question of other geometric phase arrangements such as spherical droplets.

To the Hamiltonian $\ham$ one can associate a Lagrangian $\lagr$ by inverting the sign of the potential term,
\begin{equation}
    \lagr = \frac{1}{2}\frac{d\boldsymbol\rho^{\mathsf{T}} }{dx}\boldsymbol K \frac{d\boldsymbol\rho}{dx}+V_{\rm eff}(\boldsymbol{\rho}) 
    \label{lagrangian}
\end{equation}
It is then simple to verify that, as suggested by the conservation law~\eqref{hamiltonian_z}, the corresponding Euler-Lagrange equations are identical to the equilibrium conditions~\eqref{chemPot0} (with constant $\mu_i$). The same of course applies to the Hamilton equations derived from $\ham$, once velocities $d\boldsymbol{\rho}/dx$ have been expressed in terms of momenta $\boldsymbol{p}=\partial \lagr/\partial(d\boldsymbol{\rho}/dx) = \boldsymbol{K}\,d\boldsymbol{\rho}/dx$. Finally, for a slab geometry in $d$ spatial dimensions, the interfacial free energy~\eqref{F_interface} in a system of linear size $L$ is 
\begin{equation}
\Delta F = L^{d-1}\int_0^L dx\, \lagr
\label{interface_linear}
\end{equation}
This implies that $\Delta F / L^{d-1}$ is just the action $\int dx\, \lagr$ in the mechanical analogy: it is minimized by our desired ``trajectory'' $\boldsymbol{\rho}(x)$ ($x\in[0,L]$), which physically represents the sequence of coexisting phases including the interfaces between them. As before this solution has to be chosen to satisfy the constraint of particle number conservation, which in the slab geometry reads $\int\,dx \boldsymbol{\rho}(x)=L\boldsymbol{\rho}^{(0)}$. If in addition to the slab geometry we assume periodic boundary conditions, the solution is also periodic, i.e.\ $\boldsymbol{\rho}(0)=\boldsymbol{\rho}(L)$. Finding the canonical equilibrium phase split with $n$ coexisting phases $\boldsymbol{\rho}^{(\alpha)}$ thus becomes equivalent to finding a {\em periodic Hamiltonian trajectory} visiting all $\boldsymbol{\rho}^{(\alpha)}$. Note that by considering periodic boundary conditions we are excluding the possible effects of wall interactions, another topic to which we return in the discussion at the end of the paper. 

The potential term in the Hamiltonian~\eqref{hamiltonian_z} is the {\em negative} of $V_{\rm eff}$, so that in the mechanical analogy the potential is $-V_{\rm eff}$. As a consequence, the coexisting phases are maxima rather than minima of the Hamiltonian potential. An equilibrium phase split is then a periodic Hamiltonian trajectory coming close to a number of the ``hilltops'' of this potential. Equation~\eqref{chemPot0} shows that by tuning the $K_{ij}(\boldsymbol\rho)$, i.e.\ the effective state-dependent mass tensor, one can control the shape of these trajectories $\boldsymbol \rho(\boldsymbol x)$. We will exploit this feature later in order to stabilize an interface without changing the bulk interactions.

So far we have not discussed how a system relaxes to its canonical equilibrium, as this is not our primary concern here. However, following the relaxation dynamics is in fact a convenient way to access equilibrium phase splits numerically. This dynamics is described by the Model B equation
\begin{equation}
    \dot{\rho_i} = \nabla\cdot\left [\sum_j L_{ij} \nabla\mu_j \right ]
    \label{modelB}
\end{equation}
with the chemical potentials given by~\eqref{chemPot0}. As we are not concerned with the details of the relaxation kinetics we use for the mobilities the simple form $L_{ij}=\rho_i\delta_{ij}$~\cite{Thewes2024}. We run this dynamics to steady state (see appendix for details) and use the resulting equilibrium states to confirm our theoretical predictions regarding phase splits and the effects of changing interfacial properties.

\subsection{One species plus solvent}

\begin{figure}[hbt]
\includegraphics[width=\columnwidth]{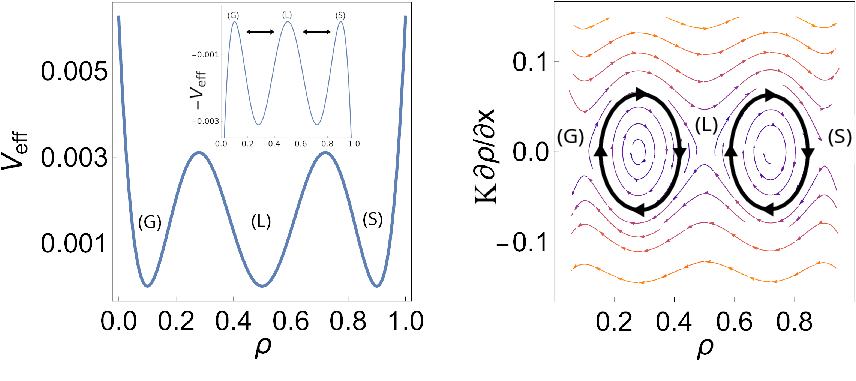}
\caption{(Left) Effective potential~\eqref{freeEnergy} with $\rho^{(\alpha)}=\{0.1,0.5,0.9\}$ and $\xi^2=0.5$. The arrows in the inset indicate the possible (bounded, periodic) Hamiltonian trajectories in the potential $-V_{\textrm{eff}}$. (Right) Flow of the 1D Hamiltonian system showing the coexistence of only two phases for any finite system. The bold lines correspond to the trajectories indicated by the arrows in the inset of the left panel.
}
\label{freeEnergy1D}
\end{figure}

We next illustrate our arguments in the simplest case of one species ($M=1$) plus solvent. We assume that, by tuning the interaction potential, a super-Gibbs number (three) of grandcanonically coexisting phases has been designed. This corresponds to an effective potential $V_{\rm eff}$ with three minima, which we represent in the simple approximate form 
\begin{equation}
    V_{\rm{eff}}(\rho) = \prod_{\alpha=1}^3 \left (1-\exp[-(\rho - \rho^{(\alpha)})^2/\xi^2] \right ).
    \label{freeEnergy}
\end{equation}
Here $\rho\equiv \rho_1$ is the density of the single molecular species present, and the $\rho^{(\alpha)}$ are the densities of the coexisting phases. The parameter $\xi$ tunes the shape of the free energy barriers between the phases; an example is depicted in Fig.~\ref{freeEnergy1D}. In a real system the functional dependence of $V_{\rm eff}$ on $\rho$ will of course be more complicated, but the above form is sufficient to capture the qualitative features. For ease of physical intuition we will denote the phases with lowest, intermediate and highest density as gas (G), liquid (L) and solid (S), respectively.

Now bearing in mind that the potential contribution in the Hamiltonian~\eqref{hamiltonian_z} is $-V_{\rm eff}(\rho)$, the only trajectories $\rho(x)$ that do not escape outside the physical range of $\rho$ (towards negative or large positive densities) are oscillations within one of the two potential wells of $-V_{\rm eff}(\rho)$ (see inset of Fig.~\ref{freeEnergy1D}). In particular, no physical trajectory $\rho(x)$ can cross the middle barrier of the potential $-V_{\rm{eff}}(\rho)$ since, by the common tangent criterion, all maxima of $-V_{\rm eff}(\rho)$ are of the same height~\footnote{This is strictly true only for an infinite system. In the Hamiltonian setup in 1d one can only use the conserved value of the Hamiltonian to change the trajectories. This allows one to change the total time, i.e.\ the system size $L$, but not the fractional volumes of the two phases. For the latter one needs to add a small tilt to $-V_{\rm eff}$ by changing the chemical potential by $\sim 1/L$. So in the generic case the maxima are not perfectly at the same height; nonetheless, their heights either decrease or increase monotonically with density so that periodic trajectories remain constrained to a single valley between adjacent maxima.}. A system with one species plus solvent therefore allows only two coexisting phases {in the canonical ensemble, even if three phases have been designed to coexist in the grandcanonical case}. Which phases will coexist is determined by the parent composition: if the parent density lies in the region between L and S (respectively G and L), these two phases will coexist in equilibrium.

\begin{figure}[bht]
\includegraphics[width=\columnwidth]{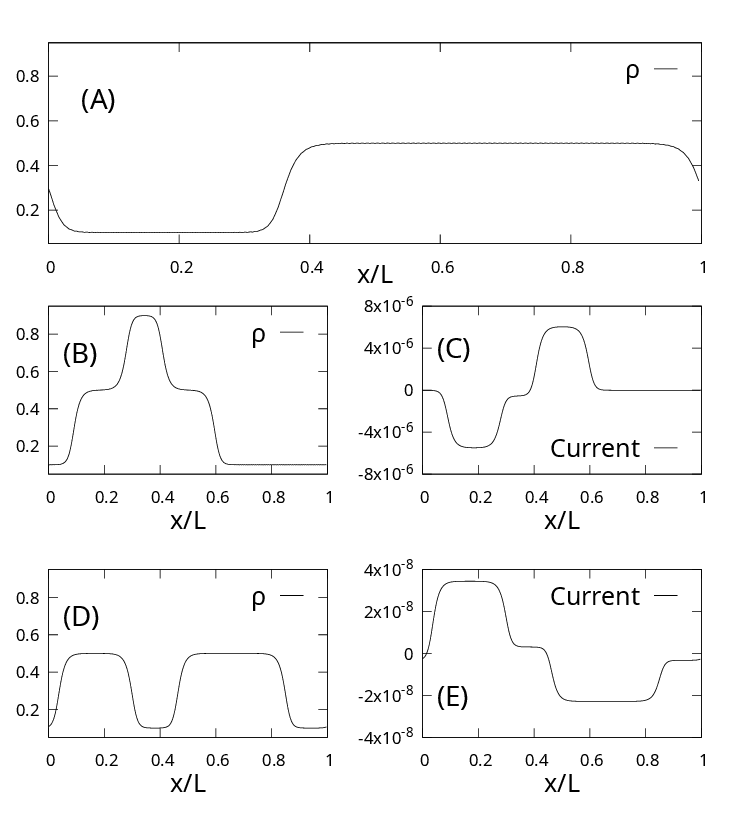}
\caption{(A) Equilibrium density profile with total density $\rho^{(0)}=0.354$ obtained via numerical integration of Eq.~\eqref{modelB}, showing coexistence of only two of the three designed phases. (B)-(E) Density and current profiles at time $t=500$ for two distinct initial conditions showing long-lived metastable states. In both cases, the eventual equilibrium profile will be as in (A). For (B,C) we use an initial condition with three domains, one for each phase. For the density profile (B) only very small currents flow as shown in (C), making this a long-lived metastable state.  For (D,E) the initial condition was uniform in space. Here, the system quickly relaxes to the two expected coexisting phases, but coarsening from two liquid domains to a single one is very slow. 
}
\label{mu1D}
\end{figure}

In order to investigate example equilibrium phase splits numerically, we solve Eq.~\eqref{modelB} in a slab
geometry of linear size $L=256$. We use a $\rho$-independent $K=3.8 \times 10^{-5}L^2$, which ensures that interface widths $\sim K^{1/2}$ are much smaller than the system size, and discretize space into intervals of length $\Delta x=L/256$. We choose initial conditions to give a parent composition $L^{-1}\int dx\, \rho(\boldsymbol x)=\rho^{(0)}<\rho^{(L)}=0.5$. In Fig.~\ref{mu1D}-(A) we show the equilibrium density profile for this set of parameters. As expected, only two of the three coexisting phases are observed, confirming our theoretical prediction.

We comment briefly on transient phenomena during the approach to equilibrium. As  shown in Fig.~\ref{mu1D}-(B) to (E), long-lived metastable states can be observed if the initial condition contains all three phases (panels B and C). The metastable nature of the configurations is evidenced by a non-vanishing but very small current flowing through the domains. The pair of interfaces between the L and S phases can then be interpreted as a ``kink'' and ``anti-kink''. Both have density profiles with exponential tails, making also their interaction decay exponentially with distance, as demonstrated by Kawasaki and Nagai~\cite{Kawasaki1983,Kawasaki1986,Kawasaki1988}. As a result coarsening is very slow, and both three-phase splits (Fig.~\ref{mu1D}-(B,C)) and multiple domains of a single phase (Fig.~\ref{mu1D}-(D,E)) can persist for long times.
Once thermal noise is introduced, we expect faster (power-law) coarsening leading more rapidly to the equilibrium state with only two coexisting phases~\cite{Kawasaki1983,Kawasaki1986,Kawasaki1988}.

We conclude that, for a system with one species plus solvent only two phases can coexist at thermal equilibrium in the canonical ensemble, even if three coexisting phases have been designed to coexist grandcanonically. This conclusion extends directly also to any larger number of designed phases, since periodic Hamiltonian trajectories will always remain confined in one of the ``valleys'' of $-V_{\rm eff}(\rho)$.

\subsection{Two species plus solvent}

The above example of a single molecular species is instructive in that it clearly shows ensemble inequivalence: even if three or more coexisting phases have been designed to coexist grandcanonically, in a canonical ensemble only two phases will appear at equilibrium, specifically those with densities either side of the density of the parent. On the other hand it is also rather special because the one-dimensional nature of the density space only allows specific phase splits: a coexistence of only gas (G) and solid (S) is impossible because a Hamiltonian trajectory (or physically, a density profile) $\rho(x)$ would always have to pass through the liquid (L) phase, too. This is no longer true once $M\geq 2$ distinct molecular species are present since $\boldsymbol{\rho}$ then lies in a higher ($M$-)dimensional composition space: here in principle any subset of the designed bulk phases can coexist without the Hamiltonian trajectory $\boldsymbol{\rho}(x)$ needing to pass through other ``intermediate'' phases. 

\begin{figure}[hbt]
\includegraphics[width=\columnwidth]{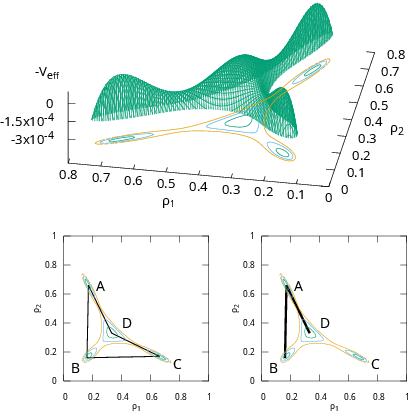}
\caption{Negative of the effective potential $V_{\rm{eff}}$ with $f$ given by~\eqref{freeEnergy2} with $\epsilon_{11}=1=\epsilon_{22}$,  $\epsilon_{12}=\epsilon_{21}=0.497804273005$ and $T=0.1806$, showing four coexisting phases (equal height maxima) in a system with two components. The bottom panels show contour lines and schematic interfaces, as lines connecting four (left) and three (right) coexisting phases.
} 
\label{2dpot}
\end{figure}

We focus in this section on the simplest case that is illustrative of the generic situation, namely, $M=2$. More specifically we use the Flory free energy (with $i,j\in \{1,2\}$)
\begin{align}
\begin{split}
    f = T\sum_i\rho_i\ln\rho_i + T\rho_0\ln\rho_0 - \frac{1}{2}\sum_{ij}\rho_i\epsilon_{ij}\rho_j
    \label{freeEnergy2}
\end{split}
\end{align}
where $\rho_0=1-\rho_1-\rho_2$ is the density of a solvent
that interacts only entropically, via volume exclusion. Fixing $T=0.1806$, $\epsilon_{11}=1=\epsilon_{22}$ and fine-tuning $\epsilon_{12}=\epsilon_{21}=0.497804273005$ results in $V_{\rm eff}(\boldsymbol{\rho})$ from~\eqref{freeEnergy2} showing four equal minima~\cite{Varea2003}, one close to each corner of the triangular region of physically allowed densities (defined by the inequalities $\rho_1,\rho_2\geq 0$, $\rho_1+\rho_2\leq 1$) and a central minimum at $\rho_1=\rho_2=\rho_0=1/3$ (Fig.~\ref{2dpot}). By our fine-tuning of $\epsilon_{12}$ we have thus obtained four phases that can coexist grandcanonically. However, in the canonical ensemble any parent composition that can be obtained as a convex linear combination of all the four phases can also be recovered with one or more subsets of only three phases, e.g.\ for the current system the three ``corner phases'' (see Fig.~\ref{sketchLever}). Adding the fourth phase to realize a super-Gibbs phase split canonically will generate additional interfaces and so can minimize the free energy only if the total interfacial free energy cost is lower than in the three-phase split. Canonical four-phase coexistence thus requires the interfacial tensions to satisfy specific conditions; it is to these that we now turn. 

To proceed, a graph-theoretic representation of the phase splits will be useful. 
Each periodic trajectory $\boldsymbol{\rho}(x)$ that is a solution of the Hamilton equations derived from the Hamiltonian~\eqref{hamiltonian_z}, representing a phase split in a large system, will visit all or a subset of the $n=4$ designed phases. Representing these phases as vertices in an abstract (fully connected) graph $G$, each phase split therefore corresponds to a closed walk on this graph. The links between nodes that are traversed on this walk correspond physically to interfaces between the corresponding phases~\footnote{The direction of the walk is unimportant here because time can be reversed in Hamiltonian dynamics or, physically, the density profiles can be mirrored, $x\to -x$.}. 

The interfacial free energy cost of any phase split is given by~\eqref{interface_linear}. In a macroscopically large system, where $L$ is much larger than all interface widths, the Hamiltonian trajectory gets exponentially close to each hilltop (bulk phase) of $-V_{\rm eff}(\boldsymbol\rho)$ that it visits, so one has a sum of contributions from trajectory segments between two phases $\alpha$ and $\beta$, which we write as $L^{d-1}\sigma_{\alpha\beta}$. The factor $\sigma_{\alpha\beta}$ is then the interfacial tension between the phases. Using that in a large system $\ham\to 0$ so that the two terms in the Lagrangian~\eqref{lagrangian} are equal~\cite{Bray1994}, 
one has explicitly
\begin{eqnarray}
\sigma_{\alpha\beta}&=&\int\,dx \sqrt{2V_{\rm eff}(\boldsymbol{\rho}) \frac{d\boldsymbol\rho^{\mathsf{T}} }{dx}\boldsymbol K \frac{d\boldsymbol\rho}{dx}}
\\
&=& \int_{{\boldsymbol\rho}^{(\alpha)}}^{{\boldsymbol\rho}^{(\beta)}} \,dl\sqrt{2V_{\rm{eff}}(\boldsymbol\rho)\frac{(d\boldsymbol{\rho}/dl)^{\mathsf{T}} \boldsymbol K (d\boldsymbol{\rho}/dl)}{(d\boldsymbol{\rho}/dl)^{\mathsf{T}}(d\boldsymbol{\rho}/dl)}}
    \label{interfTension}
\end{eqnarray}
where the second form is a line integral in composition space, with line element $dl=|d\rho|$. The denominator in~\eqref{interfTension} is then unity; we have written it explicitly to show that, compared to the single species case~\cite{Bray1994}, an effective average of the $K_{ij}$ along a direction in $\boldsymbol\rho$-space enters here~\cite{Thewes2024}. The total interfacial free energy cost of a phase split, i.e.\ a closed walk $g$ on $G$, is then
\begin{equation}
\Delta F = L^{d-1}\sigma_g, \quad
\sigma_g = \sum_{(\alpha\beta)} \sigma_{\alpha\beta}
\end{equation}
where the sum is over the edges visited. Physically, this is just the total cost of all interfaces. {In what follows, we will restrict ourselves to closed walks crossing each edge at most once in each direction. Going back and forth multiple times along an edge keeps the walk closed, but only increases $\sigma_g$ without changing the number of phases visited. We can thus safely discard such walks from our analysis}.

\begin{figure}[hbt]
\includegraphics[width=\columnwidth]{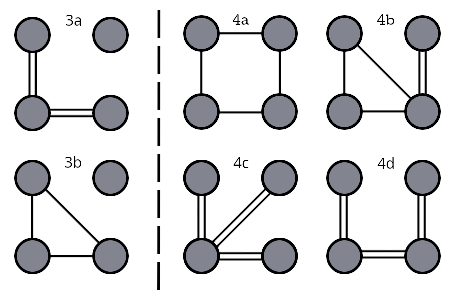}
\caption{Periodic phase splits in a system with four designed phases, represented as closed walks on a graph of four vertices. There are two (3a and 3b) topologically distinct walks representing three-phase splits, and four (4a-d) representing four-phase splits. Double lines represent edges that are crossed twice in a walk, while single lines are crossed only once.}
\label{graphs}
\end{figure}

For three-phase coexistence in a graph $G$ with four nodes, there are only two topologically distinct closed walks $g$, while for four phase coexistence there are four possibilities~\cite{Mao2019} (see Fig~\ref{graphs}). By permuting the labelling of the graph vertices by the physical phases, a number of phase splits can be generated from each of these topologically distinct closed walks. Not all of these will be able to generate a specific parent composition $\boldsymbol{\rho}^{(0)}$ via the lever rule~\eqref{lever}. However, since our goal is to find whether canonical four-phase coexistence is possible at all, we will specify the parent only at the very end, i.e.\ we ask whether there is {\em any} parent composition for which four phases will coexist.

From Fig.~\ref{graphs}, where we show the two and four topologically distinct closed walks visiting three and four vertices, respectively, we can directly rule out some candidates. Given that we have to minimize the interfacial free energy cost, topologies 4b-d will never be observed since there is always at least one double edge that can be removed to obtain one of the 3-phase topologies (3a or 3b). This latter thus always have lower total interfacial free energy, ruling out 4b-d as equilibrium phase splits. The only remaining possible four-phase topology is then 4a.

\begin{figure}[hbt]
\includegraphics[width=\columnwidth]{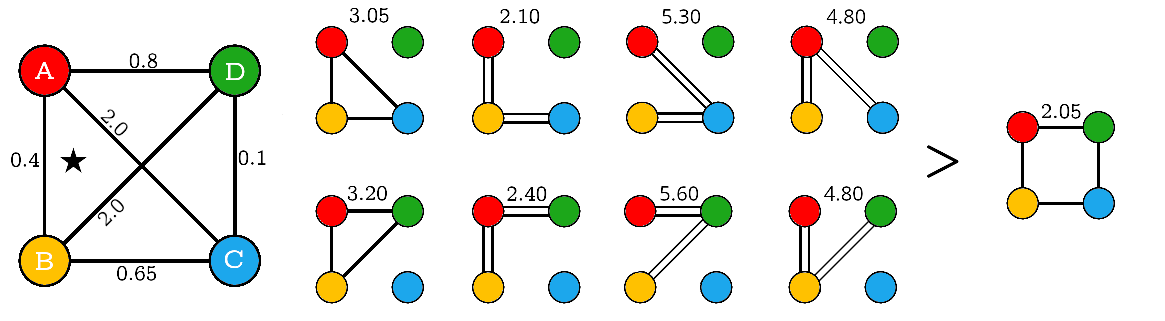}
\caption{Example of interfacial tensions (shown along the relevant graph edges) leading to a canonical equilibrium four-phase split in a system with two species plus solvent. The parent composition ($\star$) can be obtained only from two three-phase splits, ABC and ABD, and all closed walks among these phases have a higher interfacial free energy cost than the four-phase split. }
\label{4loop}
\end{figure}

We next demonstrate that it is indeed possible to design a set of interfacial tensions that satisfy $\sigma_{4a}<\sigma_{3a}$ and $\sigma_{4a}<\sigma_{3b}$, so that a four-phase canonical equilibrium phase split results. As an example, consider the case depicted in Fig.~\ref{4loop}. The topology 4a has a total cost of $\sigma_{4a}=2.05$.
Assume now that the four phases are in fact arranged geometrically as shown, in the $(\rho_1,\rho_2)$-plane and up to irrelevant linear distortions. Then we can choose a parent composition in the triangular composition space region defined as the intersection of the two triangles
ABC $\cap$ ABD. This excludes the three-phase splits ACD and BCD, as neither of these sets of phases can be linearly combined in the lever rule~\eqref{lever} to match the parent. For the remaining three-phase splits, all closed walks with topologies 3a or 3b have a higher interfacial free energy cost (see Fig.~\ref{4loop}) than the four-phase walk.

{In conclusion, the above example demonstrates that canonical four-phase coexistence with two species plus solvent is possible once we allow for designed interfacial tensions and a parent chosen within an appropriate region of composition space.}

\subsection{Multispecies case}
\label{subsec:multiSpec}
The example above can be generalized to the multispecies case, for a mixture of $M$ molecular species plus solvent. We assume as before that a super-Gibbs number $n>M+1$ of phases $\boldsymbol{\rho}^{(\alpha)}$ have been designed to coexist grandcanonically. 
Any parent composition $\boldsymbol{\rho}^{(0)}$ in the convex hull can then be 
written as a convex linear combination of these phases. But the minimum number $m$ of phases actually required in this combination is less than $n$: generically it is $m=M+1$, and it can be even smaller; e.g.\ for parents lying directly on a tieline between coexisting phases one has $m=2$. We now ask as before whether it is nonetheless possible that all $n>m$ designed phases will coexist in equilibrium in the canonical ensemble.

We again associate the $n$ grandcanonically designed coexisting phases with the $n$ vertices of a fully connected graph $G$. We assume for definiteness that the $n$ vertices are arranged in a planar fashion, as a regular polygon. We call the set of edges along the perimeter of this polygon $G^{\rm{out}}\subset G$ and denote the complementary set, containing the remaining edges, by $G^{\rm{in}}=G\backslash G^{\rm{out}}$. In the following, edges in each set will be indicated by the corresponding superscript. Since the vertices are, up to this point, unlabelled and do not correspond to the geometric arrangement of the phases in composition space, the specific labelling of the vertices around the polygon is immaterial and will become important only once we specify the parent composition.

{In a periodic slab geometry, an arrangement of coexisting phases again corresponds to a closed walk $g$ on $G$, whose interfacial free energy cost is the sum of the interfacial tensions $\sigma_{\alpha\beta}$ associated with each edge traversed by the walk. We want the cheapest -- in the sense of this free energy cost -- closed walk on $G$ to pass through all $n$ designed phases: this will guarantee that these $n$ phases coexist in a canonical setting. We now give a set of sufficient conditions on the interfacial tensions for this to be the case. 

Clearly, any closed walk on $G$ either contains a cycle, i.e.\ a closed segment on which each edge is traversed only once, or not. Let us consider the first category: any closed walk containing a cycle cannot be cheaper than that cycle. In other words, the cheapest cycle-containing closed walk must itself be a cycle, specifically the cheapest one. If this cheapest cycle runs through all $n$ phases, we can assume after appropriate relabelling that it is the cycle around the perimeter of $G$. Moreover, because $n>M+1$, the $n$-cycle always contains enough phases to recover any parent within the $n$-phase coexistence region, i.e.\ the convex hull of the $\boldsymbol{\rho}^{(\alpha)}$. 

Now the $n$-cycle on $G^{\rm{out}}$ will be cheaper than any other cycle in $G$ if and only if the cheapest path between any two vertices $(\alpha,\beta)$ lies on $G^{\rm{out}}$}, i.e.\ if
\begin{equation}
\sigma_{\alpha\beta} > \min\left[\sum_{\gamma=\alpha}^{\beta-1}\sigma_{\gamma,\gamma+1}^{\rm{out}}, \sum_{\gamma=\beta}^{\alpha+n-1}\sigma_{\gamma,\gamma+1}^{\rm{out}} 
\right]
\label{cond1}
\end{equation}
Here we have assumed that the vertices are numbered as $\gamma=1,\ldots,n$ along the $n$-cycle on $G^{\rm out}$, proceeding e.g.\ in an anticlockwise fashion. For definiteness we have also taken $\alpha<\beta$ and simplified the notation by assuming that the vertex numbering is continued periodically so that $\gamma=n+1$ is identified with $\gamma=1$ etc. The two terms being minimized over in~\eqref{cond1} are then the costs for the paths in $G^{\rm out}$ from $\alpha$ to $\beta$, in the anticlockwise and clockwise directions, respectively.

If condition~\eqref{cond1} holds for all $\alpha<\beta$ then it ensures that no internal edge, which would close any smaller cycle, will ever be crossed. Assuming one has control over all pairwise interface tensions, the set of inequalities~\eqref{cond1} can certainly be satisfied, e.g.\ by first assigning the values of the $\sigma_{\gamma,\gamma+1}^{\rm{out}}$ and then choosing appropriate values for the internal $\sigma_{\alpha\beta}$. A particularly simple such choice would be to set the interface tensions for all internal edges to some common value $\sigma^{\rm{in}} > n\langle \sigma^{\rm{out}}\rangle$.

We now turn our attention to the only remaining possibility to have fewer than $n$ coexisting phases, namely closed walks without cycles. These must be a combination of fully backtracking segments. A fully backtracking segment starts at a vertex $\alpha$, travels to a vertex $\beta$ and comes back to $\alpha$ along the reversed sequence of edges (see e.g.\ 3a and 4c-d in Fig.~\ref{graphs}). It must therefore contain an even number of edges~\cite{West2000-rg}. In addition, from~\eqref{cond1} no internal edges of $G$ can be crossed in the cheapest closed walk, and so the cheapest fully backtracking walk must be on edges contained in $G^{\rm{out}}$. Considering specifically the backtracking walk $g^{{\rm{out}},\alpha\beta}$
that travels from $\alpha$ to $\beta$ in the anticlockwise direction and back, the $n$-cycle will be cheaper if
\begin{equation}   
    \sum_{\gamma=\alpha}^{\beta-1} \sigma_{\gamma,\gamma+1}^{\textrm{out}} > \sum_{\gamma=\beta}^{\alpha+n-1} \sigma_{\gamma,\gamma+1}^{\textrm{out}}\ .
    \label{cond2}
\end{equation}
{Intuitively, backtracking from $\beta$ to $\alpha$ via the edges in $g^{{\rm{out}},\alpha\beta}$ must be more expensive than continuing in the other direction around the full cycle, along the edges not used in $g^{{\rm{out}},\alpha\beta}$. } 

The set of conditions~\eqref{cond2} is generally nontrivial to satisfy. If one fixes the parent composition then, depending on the geometry in composition space of the designed phases, multiple possible combinations of phases may satisfy the lever rule for this parent,  and the particular association of phases to graph vertices becomes important. We now investigate under which circumstances it is possible to meet the conditions~\eqref{cond2}.

\begin{figure}[hbt]
\includegraphics[width=\columnwidth]{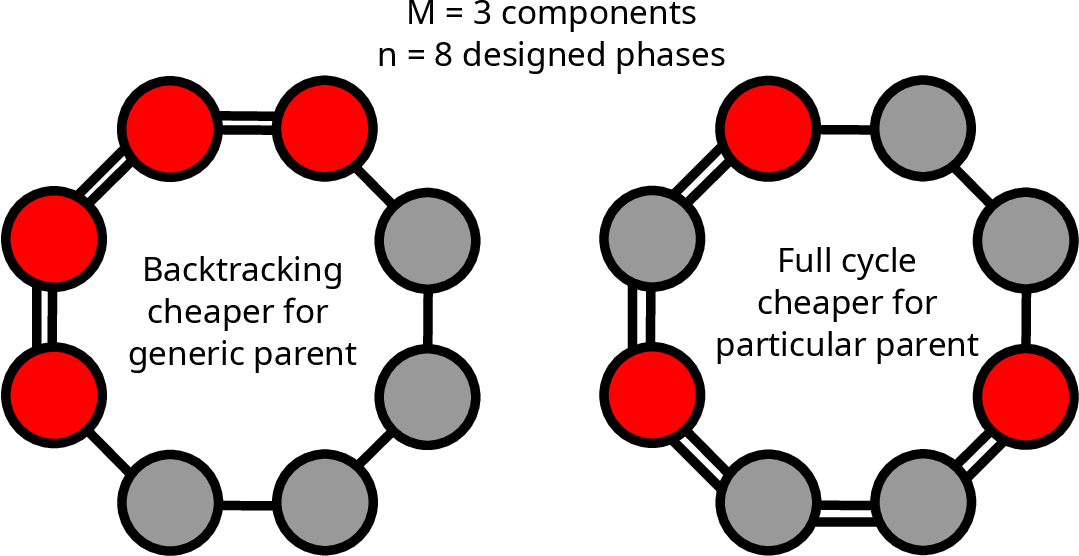}
\caption{Example with $n=8$ designed phases and $M=3$ species, with all external edges costing one unit of interfacial free energy and condition~\eqref{cond1} satisfied. For a generic parent (left), the cheapest closed walk is a backtracking walk covering $M+1$ phases, with a cost of 6. Choosing a parent such that the red phases on the right become essential allows for the eight-cycle to be cheaper (cost: 8) than the cheapest backtracking walk (cost: 10).} 
\label{5loop}
\end{figure}

First, if one needs $m$ phases to recover some \textit{fixed} parent composition according to the lever rule, a fully backtracking walk must traverse at least $2(m-1)$ edges (left panel of Fig.~\ref{5loop}), all of which must be contained in $G^{\rm{out}}$. Now set all interface tensions along $G^{\rm{out}}$ to unity, that is $\sigma_{\gamma,\gamma+1}^{\textrm{out}}=1$ for $\gamma=1,\ldots,n$. In this case, the inequality~\eqref{cond2} reduces to $m-1>n-m+1$ and if we choose a generic parent with $m=M+1$ then this is satisfied for $n < 2M$. So design of canonical super-Gibbs phase equilibria for any generic parent composition is in principle straightforward up to this number of phases by appropriate choice of the pairwise interfacial tensions.

For $n\geq 2M$, on the other hand, we can show that it is not possible to design interfacial tensions guaranteeing an $n$-phase split for a generic parent composition, at least by satisfying the sufficient conditions~\eqref{cond2}. Intuitively, this is so because the l.h.s.\ of~\eqref{cond2} contains $m-1$ terms, while the number $n-(m-1)$ of terms on the r.h.s.\ is at least as large, provided that $n\geq 2(m-1)$ (which is always true if $n\geq 2M$). To find a specific pair ($\alpha$, $\beta$) that violates~\eqref{cond2} we can then fix e.g.\ $\beta=\alpha+m-1$ and choose the node $\alpha$ that minimizes the l.h.s.\ of~\eqref{cond2}. Then the first $m-1$ terms on the r.h.s.\ are a sum of the same form and so already larger. Any parent that is a convex linear combination of the phases $\alpha,\ldots,\beta$ will then have a lower free energy cost for the $m$-phase split corresponding to the fully backtracking walk from $\alpha$ to $\beta$, compared to coexistence of all $n$ designed phases (the $n$-cycle). Overall, designing $n\geq2M$ coexisting phases thus requires restrictions on the parent composition, and this conclusion generalizes the insights from the above example for $n=4$ with $M=2$ (see Fig.~\ref{4loop}).

We next show that if one restricts the choice of parent compositions and treats it as part of the design problem, condition~\eqref{cond2} can be satisfied without an upper bound on $n$. Because the phase coexistence region in composition space is given by the convex hull of $n$ coexisting phases~\cite{Mao2019}, we can in particular consider parent compositions that are just inside the coexistence region. This fixes $M$ ``essential'' phases that span the nearest boundary of the coexistence region (a line for $M=2$, like AB in Fig.~\ref{4loop}; a plane for $M=3$ etc): these essential phases can -- together with at least one other phase -- recover the parent, and none of them can be left out. 

Let us now associate the coexisting phases to the vertices of the graph such that the $M$ essential phases are approximately equally distant from each other on $G^{\rm out}$  (see Fig.~\ref{5loop}(right)), so that for any essential phase the nearest essential phase in the clockwise or anticlockwise direction is at most $\lceil n/M\rceil$ steps away. Let also $\sigma_{\gamma,\gamma+1}^{\textrm{out}}=1$ for $\gamma=1,\ldots,n$ as before. Under these conditions, the cheapest backtracking walk covering all essential phases has a total cost of $\sigma_{\rm b}=2(n-\lceil n/M\rceil)$ while the $n$-cycle has a total cost of $\sigma_{\rm c}=n$. As a consequence, $\sigma_{\rm b}>\sigma_{\rm c}$ will always be satisfied {for $M>2$}.

Fig.~\ref{5loop} shows examples for $n=8$ designed phases and $M=3$ species. If we set all external edges to cost one unit of interfacial free energy as above, and construct the graph such that condition~\eqref{cond1} is satisfied, we cannot guarantee for a generic parent that the cheapest backtracking walk is costlier than the eight-cycle (Fig.~\ref{5loop}, left example), because $n \geq 2M=6$. However, if we are allowed to choose the parent and do this in such a way as to make the red phases in the right panel of Fig.~\ref{5loop} essential, then the eight-cycle becomes cheaper than the cheapest backtracking walk.

{While above we presented one possible route to achieving $n$ phase coexistence for $n\geq 2M$, other choices are possible once the parent composition is fixed.} In the example shown in Fig.~\ref{4loop}, the parent can be described by a linear combination of ABC or ABD (so $m=3$), but not ACD or BCD; thus A and B are essential phases. The interfacial tensions shown in the figure are not all identical along the cycle ABCDA, but nonetheless sufficient to ensure that the backtracking walks have higher interfacial free energy cost than the four-cycle. They therefore lead to an equilibrium phase split of $n=4=2M$ phases for parents with the appropriate composition.

The examples above show that it is indeed possible to achieve multiple phase coexistence beyond the standard Gibbs phase rule in the canonical ensemble, if the interfacial tensions can be arbitrarily designed to lie in appropriate ranges. (This requires in particular large interfacial tensions for the inner edges in our graph representation.)
Otherwise, even if $n$ phases have been designed to coexist grandcanonically, only a smaller number (typically $M+1$) will coexist canonically. We have thus demonstrated that, for the design of super-Gibbs phase equilibria, the grandcanonical and canonical ensembles are not equivalent: in the grandcanonical case, design of bulk interactions is sufficient, but in the canonical setup only a subset of the designed phases will generically be observed. Realizing the full number of grandcanonically designed phases canonically is possible only if {\em in addition} the interfacial tensions are designed to satisfy appropriate inequalities and, for $n\geq 2M$, if the parent composition lies in an appropriate range.

\section{Designing the Interfaces}
\label{sec:designInterf}

In this section we consider the problem of how a desired set of interfacial tensions $\sigma_{\alpha\beta}$ between designed phases could be obtained. In our continuum description, one sees from~\eqref{interfTension} that this has to be done by tuning the interfacial coefficients $K_{ij}(\boldsymbol\rho)$. Ultimately, at the microscopic level of interaction potentials $v_{ij}(\boldsymbol{r})$ between different particle species, these $K_{ij}$ could be adjusted by varying the range of the interactions, as one sees e.g.\ from a naive coarse-graining of Dean's equation~\cite{Dean1996}, which gives $K_{ij}\sim \int d\boldsymbol r \,\boldsymbol{r}^2 v_{ij}(\boldsymbol r)$; in density functional theory the interaction potential would be replaced by the direct correlation function~\cite{Hansen2006}.

One direct approach to achieving a target set of interfacial tensions would be to treat the $K_{ij}$ as optimization parameters. One would impose positive definiteness and the symmetry $K_{ij}=K_{ji}$, assuming for simplicity that $K_{ij}$ is independent of $\boldsymbol\rho$. For each matrix $\boldsymbol K$ one would then evaluate the interface profile for each pair of phases $(\alpha,\beta)$ by minimizing the free energy, and determine the interfacial tension via Eq.~\eqref{interfTension}. A loss function can then be assigned quantifying the deviation from the target values, and this is optimized numerically over the $K_{ij}$. This would be similar to the procedure used to design the bulk interactions in~\cite{Jacobs2021}.

Another possibility can be seen in Eq.~\eqref{chemPot0}. If one allows the $K_{ij}$ to depend on the composition $\boldsymbol \rho$, which is experimentally observed in colloidal systems~\cite{vis_2020} and theoretically considered in perturbation theories~\cite{Mandalaparthy2022}, the solution to Hamilton's equations can be tuned quite flexibly. Parametrizing $\boldsymbol{K}(\boldsymbol\rho)$ in such a way as to ensure positive definiteness for all $\boldsymbol\rho$ is not trivial, however. We therefore choose a representation that guarantees this automatically. In particular, consider the coordinate transformation $\boldsymbol \rho
\to {\boldsymbol{\phi}}$ with Jacobian $J_{ij}=\partial \phi_i/\partial\rho_j$. If we define a transformed effective potential in the obvious way as $V^\phi_{\rm eff}(\boldsymbol{\phi}(\boldsymbol\rho))=V_{\rm eff}(\boldsymbol\rho)$, then
\begin{equation}
     V^\phi_{\rm{eff}}({\boldsymbol\phi}) + \frac{K_0}{2}\nabla {\boldsymbol\phi}^{\mathsf{T}}\nabla{\boldsymbol\phi} = V_{\rm{eff}}({\boldsymbol\rho}) + \frac{1}{2}\nabla{\boldsymbol\rho}^{\mathsf{T}}\boldsymbol K \nabla{\boldsymbol\rho},
     \label{Map_veff}
\end{equation}
with $\boldsymbol K=K_0\boldsymbol{J}^\mathsf{T}\boldsymbol J$. This approach ensures that $\boldsymbol K(\boldsymbol\rho)$ is symmetric and positive definite for all $\bs\rho$, and maps the problem of tuning $\boldsymbol K$ to one of finding an invertible map $\boldsymbol \rho
\leftrightarrow {\boldsymbol{\phi}}$ that produces the desired interfacial tensions. {Note that because the effective potential on the r.h.s.\ of~\eqref{Map_veff} has $n$ global minima (of height zero) at the designed coexisting phases $\bs\rho^{(\alpha)}$, the transformed $V_{\rm eff}^\phi(\bs\phi)$ on the l.h.s.\ has the same structure, with the minima being at the mapped locations $\bs\phi^{(\alpha)}=\bs\phi(\bs\rho^{(\alpha)})$. Thus $V_{\rm eff}^\phi(\bs\phi)$ still represents $n$ coexisting phases. Furthermore, because of the equality~\eqref{Map_veff}, interfacial tensions calculated between these phases in the $\bs\phi$-representation are identical to those in the $\bs\rho$-representation, and the corresponding interfacial profiles are directly mapped onto each other via $\bs\phi(x)=\bs\phi(\bs\rho(x))$. To gain some intuition into the desired properties of the mapping, one notes from Eq.~\eqref{interfTension} that the interfacial tensions $\sigma$ are roughly proportional to the Euclidean distance between the phases, so that bringing two phases closer to each other in $\boldsymbol\phi$-space should decrease the cost of the interface between them.

To represent the map ${\boldsymbol \rho}\rightarrow {\boldsymbol{\phi}}$ numerically for a system with $n$ designed coexisting phases, we specify the new location $\boldsymbol\phi^{(\alpha)}$ that we want each of the original designed phases $\boldsymbol\rho^{(\alpha)}$ to be mapped to. In addition, in order to maintain the physical interpretability of the mapping, i.e.\ positive densities for all molecular species and solvent, we specify a number of points along the boundary of the physical $\boldsymbol\rho$-range (the base triangle in Fig.~\ref{2dpot}, or more generally the appropriate simplex) and insist that these are mapped to the same location in $\boldsymbol\phi$-space. We thus obtain a ``training set'' of locations ${\boldsymbol \rho}_t$ in $\boldsymbol\rho$-space and the corresponding mapped positions ${\boldsymbol \phi}_t$. We then interpolate smoothly between these by using a Gaussian Process~\cite{Rasmussen2005} for each component $i=1\ldots M$ of the mapping,
\begin{equation}
    \phi_i(\boldsymbol \rho) = C(\boldsymbol \rho, \boldsymbol \rho_t)[C(\boldsymbol \rho_t, \boldsymbol \rho_t) + \zeta^2 \mathbb{I}]^{-1}\phi_{i,t}
    \label{gaussProc}
\end{equation}
Here $C(\boldsymbol u, \boldsymbol v)$ is a kernel function;  $C(\boldsymbol \rho, \boldsymbol \rho_t)$ is the vector formed by evaluating $C(\bs\rho,\cdot)$ at all training points, $C(\boldsymbol \rho_t, \boldsymbol \rho_t)$ is the corresponding matrix resulting from evaluation for all pairs of training points, and $\phi_{i,t}$ refers to the vector of $\phi_i$-components of all target locations. To improve the smoothness of the mapping, a small regularizing term of strength $\zeta=0.01$ is added. For the kernel we choose 
\begin{equation}
    C(\boldsymbol u, \boldsymbol v) = \exp\left (-\frac{|\boldsymbol u-\boldsymbol v|}{l}\right )\left ( 1+\frac{|\boldsymbol u-\boldsymbol v|}{l} \right ),
\end{equation}
as this mimics elastic sheet behavior for large length scales $l$~\cite{Rasmussen2005} and so lends itself to the construction of invertible mappings. In the numerics below we use $l=0.4$.

We first illustrate the procedure for the effective potential~\eqref{freeEnergy} for a system with one molecular species plus solvent. Fig.~\ref{design1D} shows the mapping constructed via~\eqref{gaussProc} and the resulting $K(\rho)$ when the left (G) minimum of the effective potential is mapped from $\rho^{(1)}=0.1$ to $\phi^{(1)}=0.2$ while the other two are unchanged. Since $K(\rho)$ is now smaller between the $G$ and $L$ phases compared to its original constant value, the resulting interfacial tension between those phases decreases. 

\begin{figure}[hbt]
\includegraphics[width=\columnwidth]{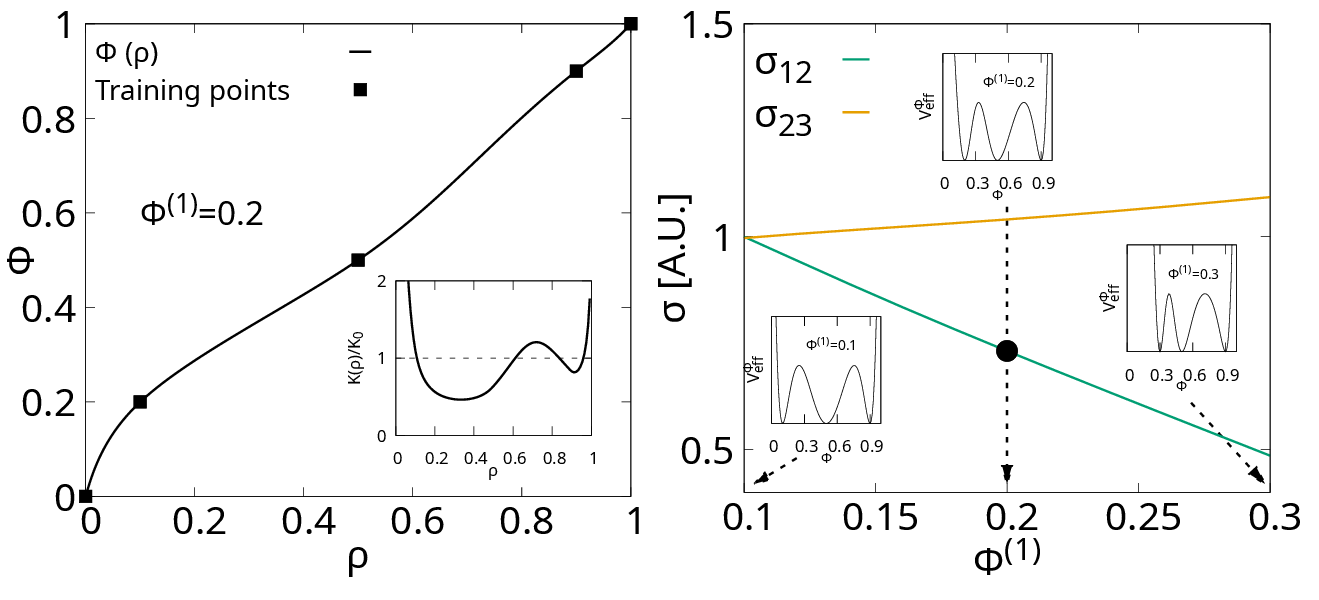}
\caption{The left panel shows the mapping $\phi(\rho)$ for training data mapping $\rho^{(1)}=0.1$ to $\phi^{(1)}=0.2$. The inset shows the resulting $K(\rho)$ (relative to the constant $K_0$ for the identity mapping) which decreases between phases $G$ and $L$, resulting in lower interface tension (black circle on the right). The right panel confirms this by showing the interfacial tension (in units of $K_0$) obtained via Eq.~\eqref{interfTension} as a function of the gas phase density $\phi^{(1)}$ for fixed $\phi^{(2)}=0.5$ and $\phi^{(3)}=0.9$, showing that bringing two phases closer together in $\phi$-space will in general decrease $\sigma$. The insets show the corresponding $V^\phi_{\textrm{eff}}$ for $\phi^{(1)}=0.1$, $0.2$ and $0.3$. } 
\label{design1D}
\end{figure}

For the single species case above we could of course have adjusted the function $K(\rho)$ directly in order to tune the interfacial tensions. We therefore next show a case where the constraint of positive definiteness is non-trivial, namely a system with two molecular species plus solvent, described by the free energy~\eqref{freeEnergy2}. We illustrate the design procedure for the challenging case of two phases that do not even form a direct interface for bland interfacial coefficients $K_{ij}=K_0\delta_{ij}$. {We first obtain numerically the location of the four minima $\boldsymbol{\rho}^{(\alpha)}$ in the grandcanonical ensemble and construct the mapping in Fig.~\ref{mapping} by keeping the location of two minima unchanged, while moving the minima at $\boldsymbol\rho \simeq (0.16,0.66)$ and $\boldsymbol\rho \simeq (0.33,0.33)$ to $\boldsymbol\phi = (0.16,0.45)$ and $\boldsymbol\phi = (0.40,0.33)$}.
\begin{figure}[hbt]
\includegraphics[trim={3.8cm 0.5cm 3.5cm 0cm}, clip, width=\columnwidth]{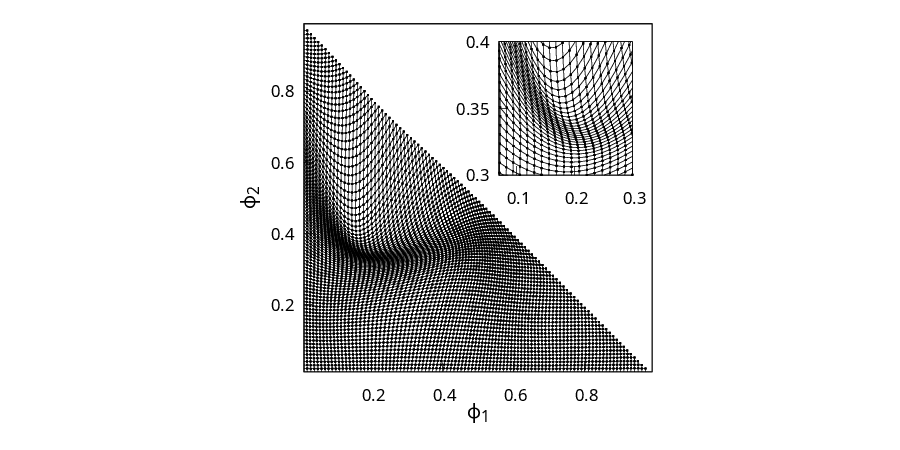}
\caption{Mapping $\boldsymbol\phi\leftrightarrow\boldsymbol\rho$. The original points lie on a regular grid in $(\rho_1,\rho_2)$ space and the points shown are the mapped values of $(\phi_1,\phi_2)$. The inset shows in detail the bijective property of the mapping, as demonstrated by the absence of  crossings of the constant $\rho_1$ and $\rho_2$ lines.} 
\label{mapping}
\end{figure}
{Finally, we perform numerical simulations of Model $B$ (Eq.~\eqref{modelB}), in the original $\boldsymbol\rho$-space,} with the resulting $\boldsymbol\rho$-dependent $\boldsymbol K=K_0\boldsymbol{J}^\mathsf{T}\boldsymbol{J}$ obtained from the Jacobian of the transformation{~\eqref{gaussProc}}. We use {$L=256$, $K_0=0.1225\times 10^{-4}L^2$ and discretize space with $\Delta x=L/512$.} 

The results in Fig.~\ref{densityProfile} show that indeed, this procedure is effective in stabilizing an interface between phases A and B that before did not exist at equilibrium. As a consequence, the interfacial design changes the phase split from BDAD, which corresponds to type 3a in Fig.~\ref{4loop} and in real space exhibits four slabs of bulk phases, to DBA, which is type 3b and only consists of three bulk slabs. Building on this proof of principle one could then exploit the full flexibility of the mapping (recall that we had fixed two of the $\boldsymbol\phi^{(\alpha)}$ 
to $\boldsymbol\rho^{(\alpha)}$ and not systematically varied the other two) to tune all interfacial tensions between the four phases, to values that will engender canonical four-phase coexistence in a two-component mixture in the canonical ensemble.

\begin{figure}[hbt]
\includegraphics[width=\columnwidth]{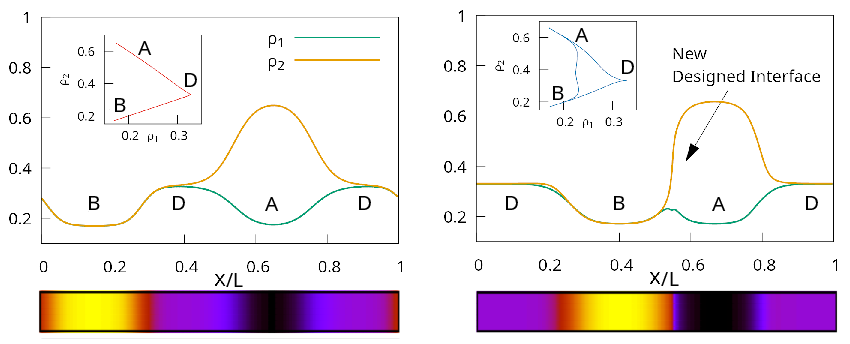}
\caption{Equilibrium density profiles with effective potential as in Fig.~\ref{2dpot} before (left) and after (right) the design of a new interface for a parent composition $\boldsymbol \rho^{(0)} = ({0.251, 0.363})$. The change in the interfacial tensions results in a new equilibrium configuration with the interface between phases $A$ and $B$ stabilized. The insets show the density profiles in the $(\rho_1,\rho_2)$ plane and illustrate the topology of interfaces. Note that the interfacial design changes the phase split from type 3a (old) in Fig.~\ref{4loop} to type 3b (new). The color plots show the phases (black, yellow, purple) arranged in a slab geometry in real space. 
} 
\label{densityProfile}
\end{figure}

\section{Discussion}

We have studied the possibility of super-Gibbs phase equilibria in $M$-component mixtures, where $n>M+1$ phases coexist. In the grandcanonical ensemble, we have given a simple derivation showing 
that fine-tuning interaction parameters introduces new degrees of freedom~\cite{Cho2023} that enter a generalized Gibbs' rule, thus enabling coexistence of more than $M+1$ phases~\cite{peters_2020,Vega1997,Akahane2016,Jacobs2021}. 

Our main focus was on understanding to what extent this results translates to the experimentally relevant canonical ensemble where particle numbers are conserved. We showed that here it is {\em interfacial properties} that determine which of the grandcanonically designed phases will actually coexist in equilibrium. This implies ensemble inequivalence for designed phase coexistence, i.e.\ coexistence in the grandcanonical ensemble does not guarantee that all coexisting phases will be observed in the canonical setup, even in the thermodynamic limit. Conversely, it is only possible to recover the same coexisting phases in the canonical ensemble if the interfacial tensions satisfy appropriate conditions.

We showed that with one component ($M=1$) in a finite box with periodic boundary conditions, coexistence of three phases in the steady state would violate conservation of an effective Hamiltonian. Thus, even though $n=3$ is possible grandcanonically, only $n=2$ equilibrium phases will appear canonically. We discussed briefly the behaviour away from equilibrium: there the situation is more complex, and metastable states containing all three phases can appear, with potentially long lifetimes when thermal noise is negligible.

For a system with two species ($M=2$) plus solvent we demonstrated, using a graph theoretical description, that a grandcanonically designed coexistence of $n=4$ phases can be realized also canonically if the interfacial tensions obey a set of inequalities, namely the typical triangle inequalities $\sigma_{\alpha\gamma} > \sigma_{\alpha\beta} + \sigma_{\beta\gamma}$ plus an additional inequality involving four phases (see Fig.~\ref{4loop}). 

The generalization to the multicomponent case consists of a combination of triangle and higher order inequalities. For the usual case where the number of phases $n$ is at most $M+1$, these inequalities simply determine the morphology of the bulk phase domains~\cite{Mao2019}, but for $n>M+1$ they become a necessary condition for coexistence in the canonical ensemble. We showed that for $n<2M$ these inequalities can be satisfied for arbitrary parent compositions within the phase coexistence region. Beyond this, we showed that assignments of the interfacial tensions can be found to realize canonical $n$-phase splits, provided that the parent composition lies close enough to a boundary of the coexistence region.

Our results thus point towards the possibility of intricate chemical engineering of phase equilibria in multi-component mixtures by tuning both bulk interactions and interfacial tensions. To move towards that goal we proposed a general method for achieving a desired set of interfacial tensions $\sigma_{\alpha\beta}$ by designing state-dependent interfacial coefficients $\bs K(\bs\rho)$. The approach exploits a tunable nonlinear transformation of composition space $\bs\rho$, to an auxiliary space $\bs\phi$ where the interfacial coefficients are constant. We illustrated the procedure in a two-component plus solvent mixture ($M=2$), showing that it can be used to stabilize a new equilibrium interface, which in turn modifies the number and spatial arrangement of the equilibrium bulk phase domains. Numerical simulations of the designed system confirmed our theoretical expectations.

We have focused in this work on phase splits that have a slab geometry, allowing for an effectively one-dimensional description. This opens an exciting avenue for future work looking at other geometries, in particular droplets. Here the Laplace pressure needs to be considered. However, at equilibrium one expects each bulk phase to occupy a finite fraction of the system. An equilibrium droplet will therefore have a radius of the order of the system size $L$ so that the Laplace pressure effects should be negligibly small. This suggests that our results should extend to phase split geometries involving droplets, though junctions between interfaces might cause additional intricacies.

Another important direction for future work will the inclusion of interactions with walls. In this paper we have used periodic boundary conditions and therefore discarded such effects, but a true canonical setup in principle requires a closed box. In one spatial dimension this should be straightforward to address, by including the two wall interactions at either end of the system as additional contributions in the interfacial free energy. In higher dimensions an additional aspect enters: even in a slab geometry, the interfacial free energy cost from interaction with the side walls (those parallel to the normal of the interfaces between bulk phase slabs) will be proportional to the fractional volume $v_\alpha$ of each phase. How this affects the conditions on interfacial tensions for super-Gibbs phase splits remains to be seen.

Overall, the possibility of designing interfaces could lead to rich phase equilibrium properties of materials, and complex functions being performed in biological environments through evolutionary engineering. 

\section*{Acknowledgments}
  This work was supported by the German Research Foundation (DFG) under grant number SO 1790/1-1. We thank Matthias Krüger for helpful discussions.

\section*{Appendix: Simulation Details}
We solve the Model B equation~\eqref{modelB} numerically in Fourier space using an implicit-explicit (IMEX) pseudo-spectral method described in~\cite{Mao2019}. Except for Fig.~\ref{mu1D}b-c, the system is always initialized with local densities $\rho_i(x)$ normally distributed around the parent density $\rho_i^{(0)}$, with standard deviation $0.01$.

In the IMEX method, the fields are updated according to
\begin{equation}
    \bs\phi(\bs q,{t+\Delta t}) = \bs\phi(\bs q,t) + \frac{\Delta\bs\phi(\bs q,{t})\Delta t}{1 + Aq^4\kappa^2\Delta t}
\end{equation}
where $\Delta\bs\phi(\bs q,t)$ is given by the spatial Fourier transform of the r.h.s.\ of~\eqref{modelB} with wave vector $\bs q$, and we use $A=0.5$ for the regularizing term. All derivatives are calculated in Fourier space while products between fields are performed in real space. The full numerical integration is performed using the Bogacki–Shampine $(3,2)$ Runge-Kutta method with adaptive time step~\cite{Bogacki1989}. An adaptive time step is especially effective in the $\boldsymbol\rho$-dependent $\boldsymbol K$ cases, where the typical interface width can vary quickly during relaxation towards equilibrium.  

\nocite{*}
%

\end{document}